\documentclass[twocolumn,floatfix,superscriptaddress,showpacs,preprintnumbers,prl,letter,aps]{revtex4}
\usepackage[dvips]{graphicx}
\usepackage{amsmath,amsbsy,amsfonts}

\usepackage{graphicx}
\usepackage{dcolumn}
\usepackage{bm}

\def\Htwoat/{H$_2$@C$_{60}$}
\def\Csixty/{C$_{60}$}
\def\Htwo/{H$_2$}
\def\para/{{\it para\/}}
\def\ortho/{{\it ortho\/}}
\def\cminv/{${\rm\,cm^{-1}}$}
\def\Kelvin/{${\rm\,K}$}
\def\Comment#1/{{\tt[#1]}}
\long\def\ToHere/{\par\Comment{**to here**}/\par}

\newcommand{\Vcoeff}[3]{\ensuremath{{{}^{#3}V_{#1}^{#2}}}}
\newcommand{\Ffun}[2]{\ensuremath{F_{#1}^{#2}}}
\newcommand{\SpherHar}[1]{\ensuremath{Y_{#1}}}

\newcommand{\Clebsch}[2]{\ensuremath{\mathcal{C}_{#1}^{#2}}}

\newcommand{\VcoefExp}[4]{\ensuremath{{{}^{#3}V_{#1}^{#2 #4}}}}
\newcommand{\A}[2]{\ensuremath{A^{#1}#2}}
\newcommand{\Abar}[1]{\ensuremath{{\rho}^{#1}}}
\newcommand{\Fbar}[2]{\ensuremath{{X}_{#1}^{#2}}}

\def\RCMvec/{\mathbf{{R}}}
\def\rvec/{\mathbf{{r}}}

\def\RCM/{R}
\def\thetaCM/{\Theta}
\def\phiCM/{\Phi}

\def\HamFiveD/{{\mathcal H}_{5D}}
\def\Evibnu/{E^{vib}_v}
\def\Erotnu/{E^{rot}_v}
\def\Bnu/{B_v}
\def\mprot/{M_p}

\def\wzeroV{\omega_0^V}
\def\wzeroT{\omega_0^T}

\def\Vnu/{V_v}
\def\Phinu/{\Phi_v}

\def\Lambdavec/{{\mathbf\Lambda}}
\def\Lvec/{{\mathbf L}}
\def\Jvec/{{\mathbf J}}
\def\LmbdaLJ/{\Lambda_{[L,J]}}

\def\cmMone/{$\mathrm {cm}^{-1}$}
\def\cmMtwo/{$\mathrm {cm}^{-2}$}

\def\Bif/{B^q_{if}}
\def\absor/{\beta}
\def\phinu/{\phi_v}
\def\Lambdaf/{\Lambda_{[L_f,J_f]}}
\def\Lambdai/{\Lambda_{[L_i,J_i]}}

\def\MJ{M_J}

\def\ML{M_L}
\def\MLambda{M_\Lambda}
\def\mlambda{m_\lambda}
\def\H{\mathcal H}
\def\HVR{\H^{VR}}
\def\HVRv{{}^v\HVR}
\def\EVR{E^{VR}}
\def\EVRvJ{{}^v\EVR_J}
\def\Bv{B_v}

\def\ket#1{\left| #1 \right>}

\def\psiT{\Psi^T}
\def\PsiT{\Psi^T}

\def\PsiV{\Psi^V}

\def\Vv{{}^v V}
\def\Vvzero{\Vv^0}
\def\Vvprime{\Vv^\prime}
\begin{document}
\title{Rotor in a Cage: Infrared Spectroscopy of an Endohedral Hydrogen-Fullerene Complex}
\author{S. Mamone}
\affiliation{School of Chemistry, Southampton University, Southampton SO17 1BJ, United Kingdom}
\author{Min Ge}
\author{D. H{\"u}vonen}
\author{U. Nagel}
\affiliation{National Institute of Chemical Physics and Biophysics, Akadeemia tee 23, 12618 Tallinn, Estonia}
\author{A. Danquigny}
\author{F.~Cuda}
\author{M. C. Grossel}
\affiliation{School of Chemistry, Southampton University, Southampton SO17 1BJ, United Kingdom}
\author{Y. Murata}
\author{K. Komatsu}
\affiliation{Institute for Chemical Research, Kyoto University, Kyoto 611-0011, Japan}
\author{M. H. Levitt}
\affiliation{School of Chemistry, Southampton University, Southampton SO17 1BJ, United Kingdom}
\author{T. R{\~o}{\~o}m}
\affiliation{National Institute of Chemical Physics and Biophysics, Akadeemia tee 23, 12618 Tallinn, Estonia}
\author{M. Carravetta}
\email{marina@soton.ac.uk}
\affiliation{School of Chemistry, Southampton University, Southampton SO17 1BJ, United Kingdom}

\date{\today}

\begin{abstract}
We report the observation of  quantized translational and rotational motion of molecular hydrogen  inside the cages of \Csixty/.  
Narrow infrared absorption lines  at the temperature of 6\Kelvin/ correspond to vibrational excitations in combination with  translational and rotational excitations and show well-resolved splittings due to the coupling between translational and rotational modes  of the endohedral \Htwo/ molecule. 
A theoretical model shows that  \Htwo/ inside \Csixty/ is  a   three-dimensional quantum rotor moving in a nearly spherical potential. 
The theory provides both the frequencies and the intensities of the observed infrared transitions. 
Good agreement with the experimental results is obtained by fitting a small number of empirical parameters to describe the confining potential, as well as the \ortho/ to \para/ ratio.
\end{abstract}


\maketitle
\newpage

Endohedral complexes of  \Htwo/ molecules  trapped inside fullerene cages have been synthesized recently\cite{Rubin_ACIE01,Komatsu_Sci05,Murata_JACS2006}.
Apart from their chemical interest and importance, these remarkable systems are ideal testbeds for the study of diatomic quantum rotors in a confined environment.
\Htwoat/ is different from quantum rotors studied so far, which were two-dimensional or showed hindered rotation, like \Htwo/ on a Cu surface\cite{Smith1996},  \Htwo/  in intercalated graphite\cite{Bengtsson2000} or C$_2$ in a metallofullerene\cite{Krause2004}.
The small mass and  large rotational constant of \Htwo/ makes it the least sensitive of molecules to the corrugations of the potential surface. 
Also, \Csixty/ provides a nearly spherical bounding potential.
Theoretically it has been shown that in \Csixty/ cages the quantum rotors CO\cite{Olthof1996}  and \Htwo/\cite{Cross_JPCA01,Xu2008HH_HD_DD} should have a measurable translation-rotation coupling, a feature that has not been experimentally resolved to our knowledge.
Moreover, there is very little experimental information on the quantum dynamics of \Htwo/ in fullerene cages\cite{Rafailov_PSS2005,Sartori_JACS2006,MC_JCP06_Ful,MC_PCCP2007}.

In general, isolated homonuclear diatomics have no infrared (IR) activity\cite{Herzberg_diatomic}. 
However, \Htwo/ does display IR activity in situations where there are intermolecular interactions present, such as in the solid and liquid phases \cite{Allen_PR1955,Hare_IR_PR1955}, in constrained environments\cite{Hourahine2003,FitzGerald_PRB02,FitzGerald_PRB06,Herman_PRB2006}, and in pressurized gasses\cite{Kudian_IR_CJP1971,McKellar_IR_CJP1974}.
IR spectra of such systems are usually broad due to inhomogeneities in the system or due to random molecular collisions.
As an exception, narrow lines are observed in semiconductor crystals\cite{Chen2002} and solid hydrogen\cite{Oka1993}.
Similarly we expect narrow lines in solid \Htwoat/, where the broadening of IR lines is suppressed by homogeneous distribution of trapping potentials provided by \Csixty/ molecules and by weak van der Waals interactions between the molecules.

In this Letter we study the dynamics of \Htwo/ in cages of \Csixty/ in the solid state with infrared spectroscopy.
The observed spectra are described by a three-dimensional quantum rotor confined in a nearly spherical potential exhibiting translation-rotation coupling.  

The \Htwoat/ powder sample (10  mg) was prepared as described in \cite{Murata_JACS2006} and pressed into  a $d=0.25$\,mm thick pellet. 
IR transmission measurements were made with an interferometer Vertex 80v (Bruker), halogen lamp, and MCT detector with an apodized resolution 0.3\,cm$ ^{-1} $. 
The sample and the reference open hole were inside an optical cryostat with KBr windows. 
The absorption coefficient $ \alpha(\omega) $ was calculated from the transmission $ T_r(\omega)$ through $ \alpha(\omega)=-d^{-1}\ln\left[ T_r(\omega)(1-R)^{-2}\right]$ with the reflection coefficient $ R=\left [ (n-1)/(n+1)\right ]^{2}$. 
A frequency-independent  index of refraction $n=2$ was assumed \cite{Homes1994}.
\begin{figure}
\includegraphics[width=0.44\textwidth]{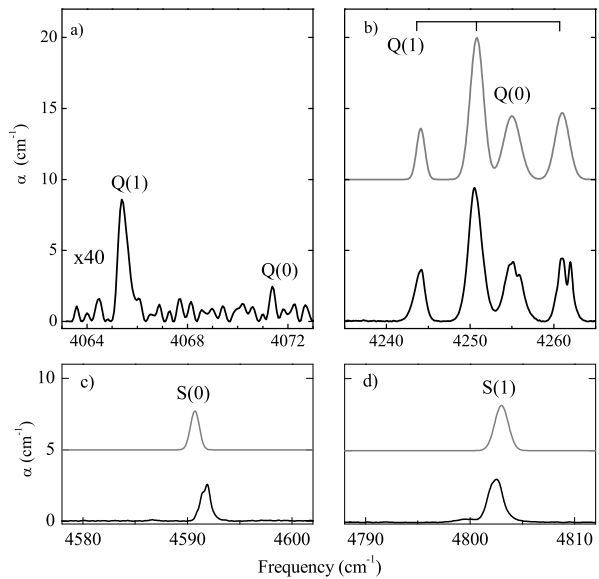}
\caption{\label{fig:experimental} Baseline-corrected  IR absorption spectra of \Htwoat/ at 6\,K  (black) and the best fit theoretical spectrum (grey) in panels (b), (c), and (d). 
All transitions are from the vibrational state $v=0$ to $v=1$. 
The  letter  indicates the change in $J$ ($Q$ for $\Delta J=0$  and $S$ for $\Delta J=2$) and the number in parentheses is the initial $J$ value. For the assignment inside a rotational branch, see Table~\ref{tab:lines} and Fig.\,\ref{fig:energylevels}. 
(a) Fundamental vibrational transitions; these are forbidden in the first-order theory so a simulated spectrum is not shown. 
(b) $\Delta J=0$, $\Delta N=+1$ transitions.  (c) \para/-\Htwo/ and (d) \ortho/-\Htwo/ $S$  transitions with $\Delta N=+1$, $\Delta J=+2$.} 
\end{figure}
The low temperature  IR  absorption peaks of \Htwoat/ are located in four narrow spectral bands between 4060 and 4810\cminv/, see Fig.\,\ref{fig:experimental}. 
This region corresponds to the \Htwo/ stretching mode and its rotational/translational sidebands. 
Peaks in 4250\cminv/, 4600\cminv/ and 4800\cminv/ regions (panels b, c, and d in Fig.\,\ref{fig:experimental}) are assigned to vibrational excitations of \Htwo/ accompanied by translational and/or rotational excitations.  
It is the translation-rotation coupling that splits the $Q(1)$ line, shown in Fig.\,\ref{fig:experimental}b, into three peaks.
Weak transitions around 4070\cminv/ (Fig.\,\ref{fig:experimental}a) represent pure vibrational excitations of the \Htwo/ molecule and are forbidden in the approximate theory presented below. 

The position and orientation of the \Htwo/ molecule is described using spherical coordinates $\RCMvec/=\{\RCM/,\thetaCM/,\phiCM/\}$  and  $\rvec/=\{r,\theta,\phi\}$ where $\RCMvec/$ is the vector from the center of the \Csixty/ cage to the center of mass of \Htwo/ and $\rvec/$ is the internuclear H-H vector.  
As first approximation we consider decoupled translational, rotational and vibrational movement of \Htwo/. 
The translation of the confined molecule may be treated using the isotropic three-dimensional harmonic oscillator model. 
The appropriate translational quantum numbers are  $N=0,1,\ldots$, the orbital angular momentum quantum number $L$, which is an integer with the same parity as $N$ and the azimuthal quantum number $\ML$. 
The radial part of the wavefunction depends both on $N$ and $L$. 
The translational eigenfunctions are
$\psiT_{NL\ML}(R,\Theta,\Phi) = \PsiT_{NL}(R) \:\SpherHar{L\ML}(\Theta,\Phi)$
where the radial wave function $\psiT_{NL}$ and the spherical harmonics \SpherHar{L\ML} are defined in \cite{FLUGGE}. 
The rotational wavefunctions, defined by the rotational quantum numbers $J=0,1,\ldots$ and $\MJ=-J,-J+1,\ldots, +J$, are given by the spherical harmonics $\SpherHar{J\MJ}(\theta,\phi)$.
It is convenient to use bipolar spherical harmonics with overall spherical rank $\Lambda$ and component $\MLambda$, defined as follows:
\begin{eqnarray}\label{eq:ffunction}
\Ffun{\Lambda\MLambda}{LJ}(\Omega)=
	\sum_{\ML,\MJ}\Clebsch{L \ML J \MJ }{\Lambda \MLambda}\SpherHar{L\ML}(\thetaCM/,\phiCM/ )\SpherHar{J\MJ}(\theta,\phi),
\end{eqnarray}
where $\Omega=\{\thetaCM/,\phiCM/, \theta,\phi\}$ and \Clebsch{}{} are  the Clebsch-Gordan coefficients \cite{VMK}. The full wavefunction describing the motion of the \Htwo/ molecule  may be written as
$\ket{vJNL \Lambda \MLambda} = \PsiV_v(r)\PsiT_{NL}(R)\Ffun{\Lambda\MLambda}{LJ}(\Omega)$
where $\PsiV_v(r)$ is the vibrational wavefunction with a quantum number $v$. 
The total nuclear spin $I$ of the \Htwo/ molecule  determines whether it is  either in a \para/ state ($I=0$ and $J$ even) or in an \ortho/ state ($I=1$ and $J$ odd).
The Hamiltonian $\H$ for the trapped molecule includes coupling terms between the vibrational, translational, and rotational motion.  
For simplicity, we neglect all   matrix elements non-diagonal in  $v$ and
 introduce a parametric dependence  on $v$:  
\begin{eqnarray}\label{eq:Ham5d}
\H &=& \HVRv +\frac{p^2}{2 m} +\Vv(R,\Omega),
\end{eqnarray}
where $\HVRv$ is the vibration-rotation Hamiltonian, $p$ is the molecular momentum operator and $m$ is the molecular mass. 
The superscript prefix $v$ is used to indicate an implied dependence on the vibrational quantum number. 
$\Vv$ is the potential energy of a molecule at a given position and orientation within the cavity, and includes terms that couple the rotational and translational motion. 
The vibrational-rotational  Hamiltonian $\HVRv$ is diagonal in the basis set $\ket{vJNL \Lambda \MLambda}$ with eigenvalues given by $\EVRvJ =\hbar\wzeroV(v+1/2)+\Bv J(J+1)$, where $\wzeroV$ is the fundamental vibration frequency;
$\Bv =B_e-\alpha_e (v+1/2)$, where $\alpha_e$ is an anharmonic correction to the rotational constant $B_e$\cite{Herzberg_diatomic}.
Below 120\Kelvin/, the thermally-activated rotational motion of the \Csixty/ cages is suppressed \cite{Tycko_PRL91,MC_PCCP2007} and $\Vv$ may be assumed to be time-independent.
Expanded in multipoles it reads:
\begin{eqnarray}
\Vv(R,\Omega) &=&\sum_{n,l,j,\lambda,\mlambda} \Vcoeff{\lambda m_\lambda}{ljn}{v} 
 R^n \Ffun{\lambda \mlambda}{lj}(\Omega),
\end{eqnarray}
where the functions $F$ are defined in Eq.~\ref{eq:ffunction} and $n$ takes even values.  
Terms with $n=2$ constitute a harmonic potential energy function, while terms with $n>2$ represent anharmonic perturbations. 
Translation-rotation coupling terms are terms with non-zero $l,j$. 

All odd-$j$ terms vanish for homonuclear diatomic molecules. 
For an icosahedral cavity, and assuming that longer-range intermolecular perturbations are negligible, all terms with odd $\lambda$ values vanish, as well as the terms with $\lambda=2$ and $4$. 
We  assume that all high-order terms starting from $\lambda=6$ are small and express the potential energy as $\Vv = \Vvzero + \Vvprime$, where the isotropic harmonic term is given by
$\Vvzero = \Vcoeff{00}{000}{v} \Ffun{00}{00} + \Vcoeff{00}{002}{v} R^2 \Ffun{00}{00}$
and the perturbation due to translation-rotation and anharmonic coupling is given by
$\Vvprime \cong \Vcoeff{00}{222}{v}R^2 \Ffun{00}{22} + \Vcoeff{00}{004}{v} R^4 \Ffun{00}{00}$.
The unperturbed Hamiltonian eigenvalues in the basis $\ket{vJNL \Lambda \MLambda}$ are given by 
$E^0_{vJNL\Lambda\MLambda }= \EVRvJ+\hbar\: {}{^v\wzeroT}(N+ 1 / 2)$,
where 
$^v \wzeroT=\left(\Vcoeff{00}{002}{v}/(2\pi m)\right)^{1/2}$ 
is the frequency for translational oscillations within the cavity. 

\begin{figure}\includegraphics[width=0.44\textwidth]{%
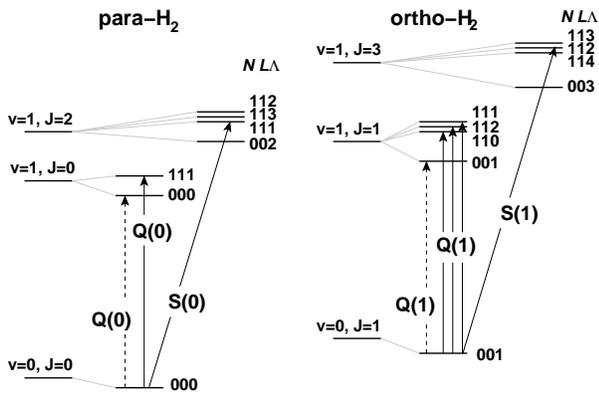}
\caption{\label{fig:energylevels}  Diagram of \Htwo/ energy levels refined against low-$T$ IR data.
For the   $v=0$ state only the ground state  rotational and translational  levels  are shown.
The  energy levels of free \Htwo/ are shown on the left while the effect of confinement by \Csixty/ is shown on the right, both for \para/- and \ortho/-H$_2$. 
The arrows show transitions corresponding to the observed low-$T$ IR  peaks. 
The transitions marked by dashed lines are forbidden within the  theory presented here.
 }\end{figure}

The matrix elements of $\Vvprime$ were evaluated analytically in the basis $\ket{vJNL \Lambda \MLambda}$ using 100 states with $N  \le 2$ and $J=1,3$  for \ortho/-\Htwo/, and  60 states with $N  \le 2$ and $J=0,2$ for \para/-\Htwo/. Matrix diagonalization leads to explicit but cumbersome expressions for the energy levels and eigenstates. 
A schematic energy level diagram is given in Fig.\ref{fig:energylevels}. The ordering of the eigenvalues depends on the relative sign and  magnitudes of the anharmonic term $\Vcoeff{00}{004}{v}$ and the translation-rotation coupling term $\Vcoeff{00}{222}{v}$. The ordering in Fig.\ref{fig:energylevels} is consistent with the experimental results.

IR activity in \Htwoat/ is due to a dipole moment, $\mu$, induced by the constraining environment.
The dipole moment operator can be expanded in multipoles depending on the instantaneous H$_2$ configuration \cite{Poll_1976}:
\begin{eqnarray}\label{eq:dipole_moment}
\mu_{q}=\frac{4\pi}{\sqrt{3}}\sum_{l,j=0}^{\infty}\A{lj}{(R,r)} \Ffun{1q}{lj}(\Omega),
\end{eqnarray}
where $q$ denotes the spherical component and the $\A{}$  coefficients describe the induced dipole moment. 
Since the dipole moment is a vector, there are restrictions on the allowed $j$ and $l$ values:
(i) $j+l$ must be odd, (ii)  $l=j\pm 1$ from the triangle relation, (iii) for homonuclear molecules only even $j$ terms are allowed. 
These restrictions imply  selection rules for IR spectroscopy of \Htwoat/. The selection rule for the total angular momentum is $\Delta \Lambda=0,\pm 1$ with the only allowed transitions having even values of $\Delta J$ and odd values of $\Delta L$. 
 In addition, because only the ground translational states ($N=L=0$) are populated at low temperature, the allowed transitions observed in the 6~K IR spectrum are to $N=L=1$ states (Fig.~\ref{fig:energylevels}).

The IR absorption amplitude at frequency $ \omega $ is \cite{Herzberg_diatomic}:
\begin{equation}\label{eq:absor}
S_ {\omega}\propto  \omega\,  \sum_{i,f}  p_i(n_K , T)  |\langle f|\mu_q|i \rangle|^2 \delta (E_{f}-E_{i}-\omega),
\end{equation}
where $K$\,=\,$O$ or $P$ selects  \ortho/- or \para/-\Htwo/. 
At the sample temperature $T$, the fractional population $p_i(n_K , T)$ of the initial state $|i\rangle$ is given by the Boltzmann distribution for \ortho/ and \para/ manifolds separately.
Since the spin isomer interconversion is negligible for the endohedral complex \cite{MC_PCCP2007}, the number of  \ortho/ and \para/ molecules, $n_O$ and $n_P$,   is not in general governed by the Boltzmann distribution and must be determined empirically.
 Since all the observed transitions are from $L=0$ to $L=1$, only terms with $l=1$ and $j=0,\:2$ are to be considered in the dipole expansion. 
This implies that only $\Delta J=0, +2$ transitions are observable.
The dipole matrix elements for the observed transitions in \Htwoat/ between states  $| i\rangle$ and $| f \rangle$ can be expressed as
\begin{eqnarray}
\langle f | \mu_{q} | i \rangle 
&=& \frac{4\pi}{\sqrt{3}} \sum_{j=0,2}   \Abar{j} [\Fbar {1q}{1 j}]_{{f} {i}},
\end{eqnarray}
where 
$\Abar{j}            = \langle \PsiV_{1}(r) \PsiT_{11}(R)|\A{1j}{(R,r)}| \PsiV_{0}(r) \PsiT_{00}(R)  \rangle$
and
$[\Fbar {1q}{1 j}]_{{f} {i}} = \langle \Ffun{f}{}(\Omega)| \Ffun{1q}{1 j}(\Omega)|\Ffun{i}{}(\Omega)\rangle $.
The matrix elements reduce into a sum of products over radial integrals $\Abar{j}$
and known angular integrals. 

Three potential parameters \VcoefExp{00}{00}{1}{2}, \VcoefExp{00}{22}{1}{2}, \VcoefExp{00}{00}{1}{4}, the rotational constant $B_e$, and $\Abar{0}$,\,$\Abar{2}$,\,$n_O/n_P$ were fitted to match the experimental frequencies and intensities. 
The potential parameters in the first vibrational state are $\{\VcoefExp{00}{00}{1}{2}, \VcoefExp{00}{22}{1}{2}\}=\{27\pm 6 , 1.5\pm 0.2\}{\rm \,J\,m^{-2}}$  and $\VcoefExp{00}{00}{1}{4} = (-2 \pm 20) 10^{20} {\rm \,J\,m^{-4}}$.
The low-$T$ data is insufficient  to derive accurately the anharmonic correction $\VcoefExp{00}{00}{1}{4}$, which is poorly defined from the separation of  $N=0$ and $N=1$  levels  only. 
This affects the value of $\VcoefExp{00}{00}{1}{2}$, because $\VcoefExp{00}{00}{1}{2}$ and $\VcoefExp{00}{00}{1}{4}$ are correlated. 
However, the translation-rotation coupling term $\VcoefExp{00}{22}{1}{2}$ is well-defined. 
 $\VcoefExp{00}{22}{1}{2}$ and $\VcoefExp{00}{00}{1}{4}$ describe the potential within approximately 0.5\AA\ from the \Csixty/ cage center, which is the root square of the average square displacement for the $N=1$ translational state.
The fitted rotational constant  is $B_e=59.3\pm 0.2$~\cmMone/ while $\alpha_e=2.98\pm 0.10 $~\cmMone/ is obtained directly from the difference in the fundamental vibrational frequencies for the \ortho/- and \para/-\Htwo/. 
The ratio between the induced dipole moment parameters is $\Abar{0}/\Abar{2}=-2.0\pm 0.2$. 
The \ortho/ to \para/ ratio $n_O/n_P=2.8\pm 0.2$ is consistent with the equilibration at any temperature warmer than 120K and suggests that there has been negligible spin isomer interconversion since the molecules were synthesized. 
The  data and the best fit results are displayed in Fig.~\ref{fig:experimental} and summarized in Table \ref{tab:lines}.
\begin{table}
\caption{\label{tab:lines}
Experimental and calculated center frequencies, $\omega$, and absorption line areas, $S_{\omega}$, of IR-active H$_2$ modes at 6\,K in \Htwoat/. 
The forbidden transitions (Fig.\,1a and dotted lines in Fig.\,2) are used as frequency references for the fitting procedure.}
\begin{tabular}{lcccccc}
\hline\hline
            & \multicolumn{2}{c}{$N L \Lambda$} & \multicolumn{2}{c}{Experimental} & \multicolumn{2}{c}{Fitted} \\
            & initial & final & $ {\omega}$ (cm$^{-1}$) & $S_{{\omega}}$ (cm$^{-2}$)  &  $ {\omega}$
(cm$^{-1}$) & $S_{ {\omega}}$ (cm$^{-2}$)     \\
\hline
$Q(1)$ & $001$ & $ 001$ & 4065.44     &  $0.093 $  &		   &	    \\
$Q(0)$ & $000$ & $ 000$ & 4071.39     &  $0.011 $  &		   &	    \\
$Q(1)$ & $001$ & $ 111$ & 4244.5        &  $5.6  $  &  4244.1	     & 4.5   \\
$Q(1)$ & $001$ & $ 112$ & 4250.7        &  $18.8 $  &  4250.7	     & 20.0    \\
$Q(1)$ & $001$ & $ 110$ & 4261.0        &  $8.7  $  &  4261.0	     & 10.0    \\
$Q(0)$ & $000$ & $ 111$ & 4255.0        &  $10.5 $  &  4255.5	     & 11.2   \\
$S(0)$ & $000$ & $ 111$ & 4591.5        &  $3.1  $  &  4590.7	     & 2.9    \\
$S(1)$ & $001$ & $ 112$ & 4802.5        &  $5.6 $  &  4803.0	     & 5.1    \\
\hline\hline
\end{tabular}
\end{table}
The frequencies of the pure vibrational transitions (Fig.~\ref{fig:experimental}a) are shifted by $-90$\,\cmMone/ from the free \Htwo/ value. 
The reduction in both the  vibrational frequency and the rotational constant \cite{Bragg1982} are consistent with a predominantly attractive C-H interaction that slightly stretches the H-H bond. 

The results described here are in qualitative agreement with previous theoretical studies. 
The quantum-chemical calculation by Cross\cite{Cross_JPCA01} gave a factor of two smaller translation-rotation coupling potential term and larger values for both harmonic and anharmonic potential terms, leading to significant discrepancies in the IR line positions if used to reproduce the experimental data.
Xu {\em et al.}\cite{Xu2008HH_HD_DD} used pair-wise Lennard-Jones potentials in their quantum-mechanical calculation.
The  ordering of the $\Lambda$ sublevels is consistent with our results, even though the splittings are different. 
Xu {\em et al.} used two different potential energy surfaces, which gave different $^v\wzeroT$, one smaller and the other larger than the experimentally observed $^v\wzeroT$.
The splitting among the $J=N=1$ energy levels, that is the measure of \VcoefExp{00}{22}{v}{2}, remained larger than the experimentally observed value for both potential energy surfaces.
Although our measurements refer to the $v=1$ excited state, while the numerical calculations are for the $v=0$ ground state, it seems that the theory using a pairwise C-H potential is not accurate enough to describe the dynamics of  \Htwoat/.
We believe that our results are a reference for theories  modeling the interaction between \Htwo/ and curved carbon nano-surfaces. 

Deformation of the cage, crystal field, or carbon isotopomers in the cage may lower the symmetry and  split the 4255 and 4261\,\cminv/ lines (Fig.~\ref{fig:experimental}b) and cause the IR activity of the weak fundamental transitions (Fig.~\ref{fig:experimental}a).

In summary, the IR spectrum of endohedral \Htwo/  displays a rich structure due to the coupled translational and rotational modes of the confined quantum rotor.
Line positions \textit{and} intensities are described by a theory involving multipole expansion of the confining potential and fitting of a small number of parameters. 
The next targets will be  analyzing higher-$T$ data  to extract information about anharmonic corrections and the vibrational ground state,  studying  lower symmetry cages and  different dihydrogen isotopomers.
The accurate determination of the energy levels in the vibrational ground state may play a key role in understanding fully the low-$T$ NMR-behavior of endohedral \Htwo/ in different fullerenes\cite{MC_PCCP2007}.

The support by the EPSRC, the EstSF grants 6138 and 7011, and the University Research Fellowship (Royal Society) is acknowledged. 
S.M.  thanks Dr. G. Pileio for useful discussions. 


\end{document}